\documentclass[aps,prl,twocolumn, showpacs]{revtex4}
\usepackage{graphicx}
\usepackage{natbib}

\begin{document}

   \title{Gravitational waves from relativistic neutron star mergers
   with microphysical equations of state}

   \author{R. Oechslin}
   \author{H.-T. Janka}
\affiliation{Max-Planck-Institut f\"ur Astrophysik, Karl-Schwarzschild-Str.~1, D-85741 Garching, Germany}

	\date{\today} 

  \begin{abstract}
        {The gravitational wave (GW) emission from a
        set of relativistic neutron star (NS) merger
        simulations is analysed and characteristic signal features
        are identified. The distinct peak in 
        the GW energy spectrum that is associated with the formation
        of a hypermassive merger remnant has a frequency that depends
        strongly on the properties of the nuclear equation of
        state (EoS) and on the total mass of the binary system,
        whereas the mass ratio and the NS spins have a weak influence.
        If the total mass can be determined from the inspiral chirp signal,
        the peak frequency of the post-merger signal is a sensitive indicator
        of the EoS.} 
   \end{abstract}
   
	\pacs{95.85.Sz, 97.60.Jd, 95.30.Lz}

   \maketitle


Among the strongest known sources of gravitational wave (GW) emission
are the merging events of double neutron star (DNS) binaries. Recent
population systhesis studies (e.g.~\citep{belczynski2001}) and
the discovery of the DNS J0737-3039~\citep{kalogera2004} suggest a
possible detection rate of GW radiation from DNS mergers
of one in $\sim$30 years for LIGO~I and one every two days for advanced
LIGO. To detect such GW signals and to filter them out of the detector 
output, theoretical waveform templates are needed.
While the inspiral phase prior to the actual merger can be described
very accurately within the post-Newtonian (PN) framework (e.g.~\cite{blanchet2002}),
hydrodynamical simulations are needed to model the dynamical merging
phase. In addition, different aspects of physics enter the
problem at this stage. Besides general relativity (GR), nuclear and
particle physics play a role in the description of the hot and dense
NS fluid via an equation of state (EoS) and in the treatment
of energy losses (e.g., by neutrinos) after the merging. The 
GW signal of the late inspiral and merging phases is 
therefore expected to contain information not only on the binary 
parameters such as masses and spins but also on the nuclear EoS.

Efforts to investigate NS mergers have concentrated
either on the relativistic aspects while simplifying the
microphysics (e.g.~\citep{shibata2006} and refs.\ therein), 
or have employed a microphysical EoS
together with an approximative neutrino treatment while
describing gravity in a Newtonian framework 
(e.g., \citep{ruffert2001, rosswog2003}). The
conformal flatness approach, a middle ground between PN and full
GR, combined with a nuclear physics-based
nonzero-temperature ($T\neq 0$) EoS has recently been chosen 
by \citet{oechslin2006b}.

\begin{figure}
\begin{center}
\includegraphics[width=7.5cm] {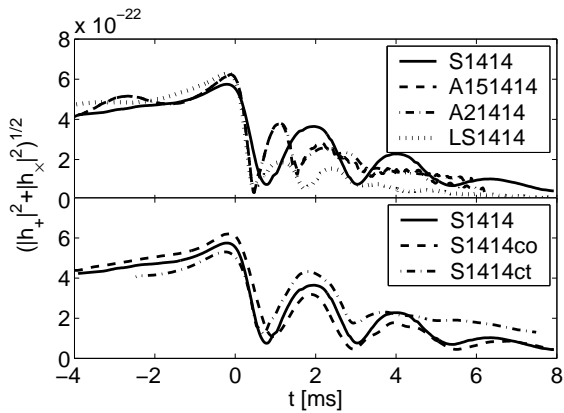}
\caption{The GW amplitude $h$ for different EoSs and
spins as radiated perpendicularly to the orbital plane of the merging
binary and measurable at a distance of 20$\,$Mpc. The minimum at about
0.5$\,$ms { is considered to mark the onset of the quasi-periodic 
wave train caused by oscillations of the rapidly spinning,
non-axisymmetric merger remnant.}}

\label{fig:gwaves}
\end{center}
\end{figure}

The generic GW signal from a NS merger can be split into a chirp-like 
part emitted by the inspiralling binary, { the burst amplitude from
the final plunge when the two stars collide (when time is set to 
$t = 0$ in Fig.~\ref{fig:gwaves}),} and a quasi-periodic 
post-merger signal caused either by the rotation and internal
oscillation of a newly formed, nonaxisymmetric hypermassive NS (HMNS) as
merger remnant, or by the
quasinormal ringing of a newly born black hole (BH) in case of a prompt
gravitational collapse of the remnant after the final plunge. {A
first maximum of the compactness of the relic HMNS is associated 
with a minimum of the amplitude $h=(|h_+|^2+|h_\times|^2)^{1/2}$ at about
0.5$\,$ms after the merging,
followed by the onset of the characteristically different, quasi-periodic 
post-merger emission.} For some of our computed models,
the quantity $h$ is plotted in Fig.~\ref{fig:gwaves}. It contains the combined
information from both polarisations $h_+$ and $h_\times$ of the wave
amplitude and therefore represents the
envelope of the high-frequency wave pattern. Its post-merger
modulation is caused by the oscillation of the nonaxisymmetric remnant.
Since the pre- and
post-merger signals are emitted in different frequency bands, they
can be clearly identified in the corresponding luminosity
spectrum. The inspiral signal leads to a broadband contribution below 
$\sim$1$\,$kHz and depends mainly on the NS
masses and their spins, while an EoS dependence is only present in the
very last stage before merging \citep{faber2002a}. On the other hand, the
post-merger signal is dominated by a quasi-periodic wave pattern with a
frequency of about 2--$4\,$kHz or about 6--$7\,$kHz, depending on whether a
HMNS forms or a prompt collapse to a BH happens~\citep{shibataprl2005}.
The associated peak in the luminosity spectrum can become very
pronounced in cases where the remnant keeps radiating
GWs for several tens of ms as suggested by recent merger
simulations \citep{oechslin2006b,shibata2006}. The bare presence of a
contribution in the frequency range of about 2--$4\,$kHz indicates the formation
of a HMNS and a nuclear EoS that is sufficiently stiff to prevent prompt
BH formation.

In the present letter, we concentrate on the HMNS formation case and
assess the question, to which extent the nuclear EoS
and the binary parameters can be constrained when such a post-merger peak
is detected. Based on a set of simulated binary merger 
models~\citep{oechslin2006b},
we identify characteristic features of the simulated GW signals and link 
them to the merger properties.
The simulations were carried out with our relativistic smoothed particle
hydrodynamics (SPH) code~\citep{oechslin2002,oechslin2004}, 
which solves the relativistic hydrodynamics equations together with 
the Einstein field
equation in the conformally flat 
approximation \citep[CFC;][]{Isenberg1980,wilson1996}. 
The simulations were started from a
stable equilibrium configuration slightly outside the innermost stable
circular orbit and the corresponding initial data were generated by
relaxing the fluid to a velocity field that includes the orbital
motion and the proper spins of the NSs. Two $T\neq 0$ EoSs,
the Shen-EoS \citep{shen1998}, and the Lattimer-Swesty-EoS
\citep{lattimer1991}, an ideal-gas EoS with parameters chosen to 
mimic the Shen-EoS, and the APR-EoS
\citep{akmal1998} were used. The APR-EoS was extended
by an ideal-gas-like thermal pressure contribution that is
proportional to the internal energy increase due to shock heating and
viscous heating \citep{shibata2006}. 
The size of this contribution is determined by an
adiabatic index $\Gamma_\mathrm{th}$ for which we chose two different values
($\Gamma_\mathrm{th} = 1.5,\,2$)
in order to investigate its influence on the merger outcome. 
Finally we calculated two models
with the Shen-EoS, restricting the latter to $T = 0$ in order to
investigate the influence of temperature-dependent pressure terms 
(see \citep{oechslin2006b}).

{ The Shen-EoS is relatively stiff and for NSs with
typical masses around 1.4$\,M_\odot$ leads to radii as big as 
$R_{\mathrm{ns}}\gtrsim$14$\,$km. Its maximum mass of non-rotating
NSs is $\sim$2.25$\,M_\odot$. In contrast, the LS-EoS is much
softer and yields a radius near 12$\,$km for a 1.4$\,M_\odot$ NS.
The APR EoS is still softer below and around nuclear
density, but becomes very stiff at higher densities ($\gtrsim
3\times 10^{14}$g$\,$cm$^{-3}$). Therefore it makes NSs even more
compact ($R_{\mathrm{ns}}\sim 11\,$km), although
it allows for a rather large maximum NS mass of 
$\sim$2.2$\,M_\odot$ compared to about
1.8$\,M_\odot$ for the LS-EoS
(see Fig.~2 in \citep{oechslin2006b}). }
Besides the EoS, we have also varied the NS masses, the mass ratio, and
the NS spins in our calculated set of models (see Table~\ref{tab:inittable}).

\begin{table}
\caption{Characteristic quantities of our computed models. 
Models with names starting with
`S' use the $T\neq 0$ Shen-EoS, `C' models the restriction 
of this EoS to $T=0$, the `LS' model uses the LS-EoS, the `P' model the 
ideal-gas EoS, and the `A15' and `A2' models the APR-EoS
extended by ideal gases with different values of 
$\Gamma_\mathrm{th}$. 
All models were computed with irrotating initial conditions except
the last four cases where the ending `co' (`ct') of the model names
indicates initially corotating (counterrotating) spin states
of the NSs (for the spin frequencies, see \citep{oechslin2006b}).
$M_1$ and $M_2$
are the individual gravitational masses in isolation, and
$q=M_1/M_2$ is the mass ratio.
$f_\mathrm{max}$, $f_\mathrm{peak}$, $\Delta E_\mathrm{in}$, and
$\Delta E_\mathrm{pm}$ are defined in the text. `SNR' means the estimated
signal-to-noise ratios in advanced LIGO (left) and DUAL (right) for the
GW emission after merging and a source distance of 20$\,$Mpc. 
}
\begin{tabular}{c|c|c|c|c|c|c|c|c|c}
Model & $M_1$ & $M_2$ & $q$ &
$f_\mathrm{peak}$&$f_\mathrm{max}$&$\Delta E_\mathrm{in}$&$\Delta E_\mathrm{pm}$&\multicolumn{2}{|c}{$\mathrm{SNR}$}\\
\hline
 & $M_\odot$ & $M_\odot$ & &kHz&kHz&\multicolumn{2}{|c|}{$10^{-3}M_\odot$}&\\
\hline
S1414 & 1.4 & 1.4 & 1.0 &2.24&1.31&6.2&5.3&2.7&3.2\\
S135145 & 1.35 & 1.45 & 0.93 &2.27&1.35&6.2&6.4&2.8&3.3\\
S1315 & 1.3 & 1.5 & 0.87 &2.26&1.29&5.5&6.6&2.7&3.3\\
S1216 & 1.2 & 1.6 & 0.75&2.20&1.18&4.4&4.1&2.2&2.6\\
S1515 & 1.5 & 1.5 & 1.0&2.45&1.45&8.4&8.4&2.8&3.6\\
S1416 & 1.4 & 1.6 & 0.88 &2.37&1.32&7.8&9.4&3.1&3.8\\
S1317 & 1.3 & 1.7 & 0.76&2.39&1.20&6.4&6.4&2.4&3.1\\
S1313 & 1.3 & 1.3 & 1.0&2.16&1.39&4.6&4.0&2.6&2.9\\
S1214 & 1.2 & 1.4 & 0.86&2.08&1.24&4.1&3.8&2.5&2.8\\
S1115 & 1.1 & 1.5& 0.73&2.10&1.10&4.2&3.8&1.8&2.2\\
\hline
C1216 & 1.2 & 1.6 & 0.75 &2.34&1.19&4.5&3.4&1.9&2.4\\
C1315 & 1.3 & 1.5 & 0.87 &2.37&1.27&5.6&6.1&2.6&3.2\\
P1315 & 1.3 & 1.5 & 0.87 &2.13&1.28&5.9&4.0&2.4&2.8\\
\hline
LS1414 & 1.4 & 1.4 & 1.0 &3.67&1.81&11.1&2.5&1.0&1.4\\
A151414 & 1.4 & 1.4 & 1.0 &3.63&1.90&15.3&20.0&2.1&3.7\\
A21414 & 1.4 & 1.4 & 1.0 &3.45&1.90&16.0&19.2&2.2&3.8\\
\hline
S1414co & 1.4 & 1.4 & 1.0 &2.28&1.47&7.7&3.4&1.9&2.5\\
S1414ct & 1.4 & 1.4 & 1.0  &2.24&1.19&5.3&9.3&3.0&4.1\\
S1216co & 1.2 & 1.6 & 0.75 &2.23&1.13&4.6&0.6&0.8&1.0\\
S1216ct & 1.2 & 1.6 & 0.75 &2.14&1.11&3.6&5.7&2.7&3.3\\
\end{tabular}
\label{tab:inittable}
\end{table}


\begin{figure}
\begin{center}
\includegraphics[width=7.5cm]{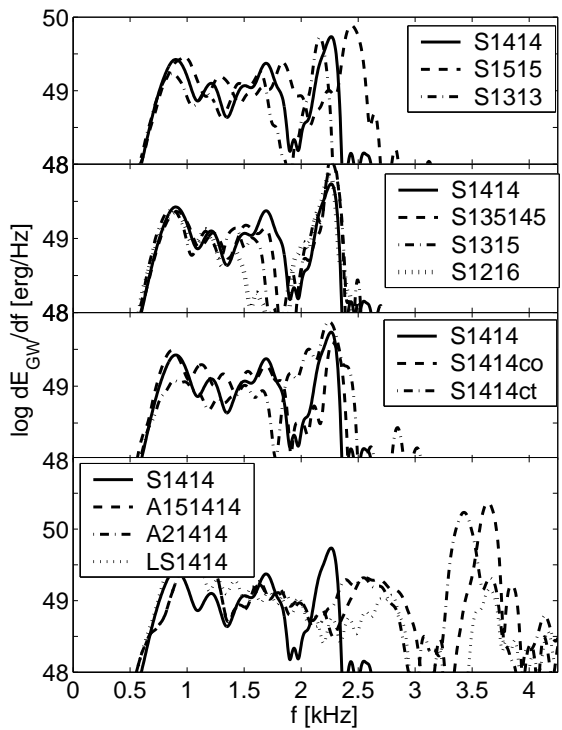}
\caption{GW luminosity spectra for different total masses, mass
ratios, spin configurations, and EoSs (from top to bottom). Note that the
spectrum below 1$\,$kHz is not represented correctly because we started
the simulations when the GW frequency was $\sim\,$1$\,$kHz.}
\label{fig:luminosityspectra}
\end{center}
\end{figure}

The GW waveform $h_{ij}$ is extracted by making use of the 
quadrupole formula and
is given by $h_{ij}=(2/D)d^2Q_{ij}/dt^2$, where $Q_{ij}$ is the 
Newtonian mass quadrupole
and $D$ is the distance from the source (the indices $i$,$j$ denote the
spatial directions). 
Compared to a more detailed extraction technique in the wave
zone using the gauge-invariant Moncrief variables as done, e.g., in
\citep{shibata2006}, this approximation is able to describe the GW
signal only qualitatively. The wave phase and thus the frequency
information can be well reproduced but the amplitudes are
underestimated by about $30\%$ in the inspiral regime and by about 
$40\%$ in the post-merger regime \citep[][Fig.~12, panel a]{shibata2005}.
Based on the thus obtained waveform, the GW luminosity spectrum 
can be determined according to \citep{Thorne300years} by
$dE_\mathrm{GW}/df = 
\frac{\pi}{2}4\pi D^2 f^2\langle|\tilde{h}_+|^2+|\tilde{h}_\times|^2\rangle$,
where $\tilde h_+$ and $\tilde h_\times$
 denote the fourier transforms of the
waveforms of `$+$' and `$\times$' polarisation, respectively.
The angle brackets indicate averaging over all possible source detection
angles. The energy emitted in GWs is then given by
$\Delta E_\mathrm{GW}=\int df\, dE_\mathrm{GW}/df$.
Because of the underestimation of the GW amplitude, the GW luminosity
spectra, which depend quadratically on the amplitude, are 
systematically too low by $\sim$70\%. 

Bearing this in mind, we consider in the following quantities
that are not directly affected by this shortcoming and independent
of the source orientation, namely 
(see Table~\ref{tab:inittable}):
(i) $f_\mathrm{max}$ as the frequency of the GW signal
when the amplitude becomes
maximal at about the time of the final plunge.
It is determined by fitting a function
of the form $A(t)\cos(\omega(t)t+\phi)$ to the waveform;
(ii) $f_\mathrm{peak}$ as the frequency of the post-merger peak in the
luminosity spectrum;
(iii) the ratio $\Delta E_\mathrm{in}/\Delta E_\mathrm{pm}$ of $\Delta
E_\mathrm{in}$ as the energy emitted over a time interval of 3$\,$ms
before merging, and $\Delta E_\mathrm{pm}$ as the energy radiated over
a time interval of 5$\,$ms after merging.
The energies are determined as described above
from the waveforms produced in the corresponding time intervals.
The values thus obtained agree with time integrals of the
quadrupole-formula-based expression $dE/dt=1/5 \langle\dddot Q_\mathrm{ij}\dddot Q_\mathrm{ij}\rangle$
to within $\sim$20\%.

\begin{figure}
\begin{center}
\includegraphics[width=7.5cm]{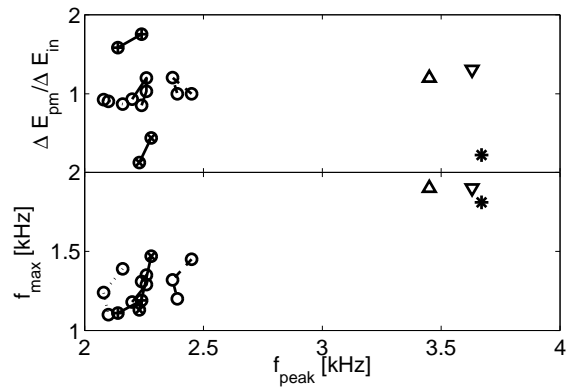}
\caption{Ratio $\Delta E_\mathrm{pm}/\Delta E_\mathrm{in}$
vs.\ $f_\mathrm{peak}$ (top) and $f_\mathrm{max}$ vs.\ $f_\mathrm{peak}$ (bottom) for the models
considered. Shen-EoS models are shown with a circle, APR-EoS models
with triangles, and the LS-EoS model with a star. 
The two corotating models are indicated by `$\times$', while
the counterrotating models are marked with `$+$', The horizontal
spread of the model group with the Shen-EoS is mainly caused by different
total system masses. Lines connect cases with the same total 
mass and spin setup.}
\label{fig:summary}
\end{center}
\end{figure}

In cases where a HMNS forms, $f_\mathrm{peak}$ turns out to depend 
sensitively on the
EoS (Fig.~\ref{fig:luminosityspectra}, bottom) and to a lesser extent
on the total mass of the binary system 
(Fig.~\ref{fig:luminosityspectra}, top). The NS spins and the
mass ratio have very little influence
(Fig.~\ref{fig:luminosityspectra}, middle panels). Indeed, all models
using the Shen-EoS lead to values around 2.1--2.4$\,$kHz for
$f_\mathrm{peak}$, where the variation among the
Shen-models of about 0.3$\,$kHz is mostly due to the total system mass. On the
other hand, the models using the APR-EoS
and LS-EoS with their more compact NSs do not only yield larger values for
$f_{\mathrm{max}}$ \citep{shibataprl2005,faber2002a} but also much larger ones 
(around 3.6$\,$kHz) for $f_\mathrm{peak}$. { The latter
quantity mainly depends on the behavior of the EoS in the 
density regime between 0.5$\rho_0$ and 2$\rho_0$ ($\rho_0 \approx
2.7\times 10^{14}\,$g$\,$cm$^{-3}$ being the nuclear saturation density),
where the bulk of the remnant mass is located. While peak 
temperatures of several 10$\,$MeV are present in the HMNS
at such densities (see \cite{oechslin2006b}), $T\neq 0$ 
contributions to the gas pressure affect the basic properties of 
the post-merger oscillations only moderately. A
comparison of models C1315, C1216, and A151414 with S1315, S1216, and 
A21414, respectively, shows that $f_\mathrm{peak}$ decreases
by at most $\sim$0.2$\,$kHz when thermal pressure is included 
(Table~\ref{tab:inittable}). This is caused by the less compact 
structure of the HMNS in these cases. }
Considering the radiated energies before and after merging, 
$\Delta E_\mathrm{in}$
and $\Delta E_\mathrm{pm}$, respectively, we find a
characteristic variation of the GW signal with the NS spins. 
As shown in Fig.~\ref{fig:summary}, the ratio
of $\Delta E_\mathrm{pm}$ to $\Delta E_\mathrm{in}$ is highest for
counterrotating cases and lowest for corotating NSs. This is so because
corotation leads to a stronger inspiral signal due to a positive
contribution from the NS spins, while damping the amplitude of the
post-merger part due to a smaller non-axisymmetry of the remnant
\citep{oechslin2006b}. Counterrotation has the opposite effect 
(cf.\ Fig.~\ref{fig:gwaves}).
The degeneracy of $f_{\mathrm{max}}$ visible in Fig.~\ref{fig:summary} 
for cases with APR and LS EoS can be lifted when the ratio 
$\Delta E_\mathrm{pm}/\Delta E_\mathrm{in}$ is taken into account.

\begin{figure}
\begin{center}
\includegraphics[width=8cm]{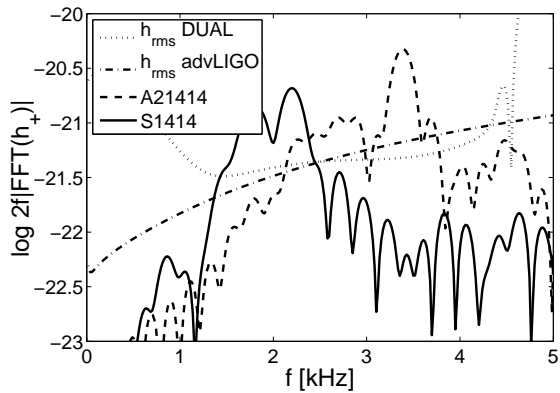}
\caption{GW spectra $2f|\tilde h_+(f)|$ for two typical models with
different EoSs, assuming a source distance of 20$\,$Mpc. Also shown is the
strain noise amplitude $h_\mathrm{rms}$ of advanced LIGO and of the
resonant spheres detector DUAL. }
\label{fig:hchar}
\end{center}
\end{figure}


{ A NS merger therefore produces a GW signal whose
location in the space of the parameters $f_\mathrm{max}$,
$f_\mathrm{peak}$, and $\Delta E_\mathrm{pm}/\Delta E_\mathrm{in}$
depends distinctively on the properties of the nuclear EoS.
The latter determines the compactness of the merging stars
and of the HMNS and thus the GW frequencies and energies
emitted during the final plunge and post-merger oscillations. 
Characterizing a GW measurement in terms of the
three parameters therefore provides direct information about the
NS EoS, in particular if the system mass is known from the
inspiral chirp signal. This is a promising alternative to constraining
the EoS by NS mass and very difficult 
radius determinations (e.g., \cite{LattPrak07}). More work, however,
is needed to understand the post-merger oscillations in terms of
involved eigenmodes of the HMNS (\cite{Kokkotas2001} and refs.
therein), and the GW parameters for a large
variety of EoSs should be computed by accurate GR merger 
simulations. }

{ To assess the detectability of the post-merger GW emission we follow 
Ref.~\citep{flanagan1998} and consider the advanced LIGO
interferometer and the omnidirectional DUAL detector, which consists of
two nested resonant spheres \cite{Cerdonio2001}. 
The signal-to-noise ratio (SNR) for
a given GW signal $h(t)$ can be written as
$\mathrm{SNR}^2=\int_{-\infty}^\infty d\ln f|2f\tilde h(f)|^2/h_\mathrm{rms}(f)$, 
where
$\tilde h(f)$ is the fourier transform of $h(t)$ and
$h_\mathrm{rms}(f)$ is the strain noise of the detector. 
In Fig.~\ref{fig:hchar}, we take $h(t)=h_+(t)$ and compare the spectrum
$2f|\tilde h_+|$ for two representative models with the strain noise
of the LIGO and DUAL instruments. For an
interferometer, $h$ depends on the source orientation and direction
relative to the interferometer arms and in the optimally aligned case
is equal to $h_+$ or $h_\times$. For DUAL, $|\tilde h(f)|^2=|\tilde
h(f)_+|^2+|\tilde h(f)_\times|^2$, because both
polarisations can be measured simultaneously. }
In Table
\ref{tab:inittable} the SNRs are listed for all of our models. 
Note that only the post-merger waveforms
are considered and a distance of 20$\,$Mpc is assumed. 
Since we have underestimated the wave amplitudes
by $\sim$40\% (see above), our SNRs
may be too low by up to a factor of 1.7. Moreover, the post-merger
signals are likely to be emitted for longer times than
the considered window of 5$\,$ms. Fitting an exponential
decay to the GW amplitudes, we find a decay time
of about 5$\,$ms. From this, a further increase of the SNR by
$\sim$15\% is estimated. Taking these corrections into account,
we obtain for typical models like S1414 and APR21414 a
SNR of { $\sim$5 in LIGO and $\sim$6.5 in DUAL. 
Assuming that a minimal value of about 3 is needed for detection
in case the preceding inspiral chirp has been measured, such GW signals
may be identified up to $\sim$35$\,$Mpc (LIGO) and
$\sim$45$\,$Mpc (DUAL).
According to \cite{kalogera2004,kim2003},
these distances correspond to event rates of 0.04--0.5
(LIGO) and 0.08--1.1 (DUAL) per year. }
\begin{acknowledgments}
We thank the DUAL research team for advice. This work was supported by the DFG through grants SFB-TR~7 
and SFB-375.

\end{acknowledgments}

\bibliography{../discpaper/biblio}
\end{document}